\documentclass[prl,aps,twocolumn]{revtex4}
\usepackage{amssymb}

\begin{document}

\title{Reply to the Comment by Galanakis \textit{et al} on the paper
\textquotedblleft Exact bosonization for an interacting Fermi gas in
arbitrary dimensions" }
\author{K. B. Efetov$^{1}$,C. P\'epin$^{2}$, H. Meier$^{2}$ }
\affiliation{$^{1}$Theoretische Physik III, Ruhr-Universit\"{a}t
Bochum, 44780 Bochum, Germany} \affiliation{$^{2}$IPhT,
CEA-Saclay, L'Orme des Merisiers, 91191 Gif-sur-Yvette, France}

\begin{abstract}
It is shown that the criticism presented in the Comment by
Galanakis et al \cite{1} on the paper by Efetov et al \cite{2} is
irrelevant to the bosonization approach.
\end{abstract}

\maketitle
\date{\today}

In spite of the far going conclusion that our bosonization scheme
cannot help to overcome the negative sign problem, the Comment
does not really address the suitability of the method for Monte
Carlo (MC) calculations. We wrote in the paper that our mapping of
the fermionic model onto a bosonic one was exact having in mind
the continuous with respect to (imaginary) time limit. Any quantum
MC scheme implies generically a discrete time with extremely
strong variation of the Hubbard-Stratonovich (HS) field $\phi $ on
different slices. In this case, our bosonization scheme is not
necessarily exact for a $\emph{given}$ HS field and we make an
approximation. If everything remained exact also in this case, we
would not be able to overcome the sign
problem because the partition function $Z\left[ \phi \right] $ for a given $%
\phi $ would be the same for both bosonic and fermionic representations and
could be negative.

In our opinion, the comment of Galanakis \textit{et al.} is based
on a misunderstanding of the definition of $Z_{b}$. Indeed $Z_{b}$
should not be understood as coming from Eqn (9) but from
Eqns(12-13). [We apologize for this confusion, which is the result
of writing the paper several times in order to improve the
presentation]. Having failed to understand this point, Galanakis
\textit{et al.} re-derive the same steps leading to $Z_{f}$ and
jump on the far reaching conclusion that our technique does not
solve the MC sign problem. However, the bosonization method starts
later with Eqs. (12, 13), which means that Galanakis \textit{et al
}make their conclusions not about the bosonized model but still
about the original fermionic one.

They fail as well to understand that for $\emph{finite }$ time
slices, our method is not exact but requires an approximation. The
crucial approximation is made when writing Eq. (13). According to
Eq. (11), the function $A_{r,r^{\prime }}\left( \tau \right) $ is
expressed in terms of the Green functions at slightly different
times $\tau $ and $\tau +\delta ,$where $\delta
\rightarrow +0$. Therefore, deriving the equation for the function $%
A_{r,r^{\prime }}$ and making no approximations one would have the fields $%
\phi _{r}$ and $\phi _{r^{\prime }}$ at slightly different times $\tau $ and
$\tau +\delta $. Nevertheless, we put $\delta =0$ in Eq. (13). In the
continuous limit, (implying subsequent averaging over $\phi _{r}\left( \tau
\right) $), this approximation becomes exact. Therefore our field theory
based on the introduction of superfields is exact.

However, working with finite slices of time and putting $\delta =0$ changes
the function $Z_{b}\left[ \phi \right] $ for a given configuration of $\phi $%
. Taking Eq. (13) with the fields $\phi _{r}\left( \tau \right) $ and $\phi
_{r^{\prime }}\left( \tau \right) $ at coinciding times $\tau $ results in a
symmetry of the solution $A_{r,r^{\prime }}\left( \tau \right) $ under the
replacement $r\leftrightarrows r^{\prime }$. We argue in the paragraph after
Eq. (23) that any possible singularity in the integral over $u$ in the
exponent should be absent due to this symmetry. This means that the
imaginary part in the exponent does not arise and the function $Z_{b}\left[
\phi \right] $ must be positive.

In order to make everything well defined one needs a regularization,
otherwise the solution of Eq. (13) is not uniquely defined. [The fact that
the solution of Eq. (13) is not uniquely defined is actually remarked in the
Comment in the paragraph before the conclusion]. In our PRL paper we had no
possibility to discuss this question due to the lack of the space. However,
we emphasize that Eq. (13) itself and the way we choose its solutions lead
to a real exponent in Eq. (12) and to a positive $Z_{b}\left[ \phi \right] $%
, thus providing a method of calculations free of the sign problem.

The regularization of Eq. (13) is absolutely necessary before this
equation can be used for explicit numerical computation. This goal
is achieved in our recent preprint \cite{3} where all details of
the derivation of Eq. (13) are discussed and an explicit
regularization is introduced. We make a check of our scheme by
explicit diagrammatic expansions and demonstrate perfect
agreement. We give detailed explanations about the role of the
symmetry $r\leftrightarrows r^{\prime }$ for removing the
singularity of regularized solution of Eqn (13). We present there
an explicit formula, Eq. (4.49), that can directly be used for
numerics. We have checked it analytically and numerically for a
static $\phi _{r}$. This formula gives by construction real
positive $Z_{b}\left[ \phi \right] $ for $\emph{any}$ time
dependent fields $\phi _{r}\left( \tau \right) $, property which
should make the computation free of the sign problem.

We conclude that the claim of the Comment that our method cannot
resolve the sign problem is not correct. Moreover, it is not based
on a discussion of the bosonization scheme. The authors of the
comment have repeated the derivation of the equations preceding
the key equation (13) of \cite{1} where the true bosonization
starts. Although their discussion helps to correct several
inaccuracies in the PRL paper, they miss the point that a crucial
approximation is made for the piece-wise fields $\phi _{r}\left(
\tau \right) $ used in the MC calculations and thus come to an
incorrect conclusion. As it stands, the Comment of Galanakis
\textit{et al.} is not scientifically valid.

\end{document}